\documentclass[11pt]{article}
\usepackage{graphicx}
\usepackage[margin=2.25cm]{geometry}%
\usepackage{bm}

\def\prl#1#2#3{{\it Phys. Rev. Lett.} {\bf #1}, {\it #2}, #3.}
\def\jcompphys#1#2#3{{\it J. Comp. Phys.} {\bf #1}, #2 (#3).}

\def\molsim#1#2#3{{\it Mol. Simulat.} {\bf #1}, {\it #2}, #3.}

\def\arpc#1#2#3{{\it Ann. Rev. Phys. Chem.} {\bf #1}, {\it #2}, #3.}

\def\jcp#1#2#3{{\it J. Chem. Phys.} {\bf #1}, {\it #2}, #3.} 
\def\jpc#1#2#3{{\it J. Phys. Chem} {\bf #1}, {\it #2}, #3.}
\def\jpcb#1#2#3{{\it J. Phys. Chem. B} {\bf #1}, {\it #2}, #3.}

\def\cpl#1#2#3{{\it Chem. Phys. Lett.} {\bf #1}, {\it #2}, #3.}

\def\chemrev#1#2#3{{\it Chem. Rev.} {\bf #1}, {\it #2}, #3.}

\def\mp#1#2#3{ {\it Mol. Phys.} {\bf #1}, {\it #2}, #3.}

\def\science#1#2#3{{\it Science} {\bf #1}, {\it #2}, #3.}

\def\jpcm#1#2#3{ {\it J. Phys.: Condensed Matter } {\bf #1}, {\it #2}, #3.}

\def\expt#1{\langle #1\rangle}
\def\be{\begin{equation}}
\def\ee{\end{equation}}

\begin{document}
\baselineskip 20pt
\begin{center}

{\Large \bf Multiple Time-Scale Behaviour and Network Dynamics in Liquid Methanol}\\
\qquad \\
{\bf Ruchi Sharma} and 
{\bf Charusita  Chakravarty}$^*$\\
Department of Chemistry, \\
Indian Institute of Technology-Delhi,\\
New Delhi: 110016, India.\\
{\bf Edoardo Milotti}$^{\dag}$\\
Dipartimento di Fisica, \\
Universit\`a di Trieste and I.N.F.N. -- Sezione di Trieste,\\
Via Valerio 2,  I-34127 Trieste, Italy.\\
 \quad\\
 {\bf Abstract}
 \end{center}

 Canonical ensemble molecular dynamics simulations of
 liquid methanol, modeled using a rigid-body, pair-additive
 potential, are used to compute  static distributions and temporal 
 correlations of tagged molecule potential energies as a means
 of characterising the liquid state dynamics. 
The static distribution of tagged molecule potential energies shows a
clear multimodal structure with three distinct peaks, similar to those
observed previously  in  water and liquid silica.  The multimodality
is shown to originate from electrostatic effects, but not from local,
hydrogen-bond interactions.
An interesting outcome of this study is the remarkable similarity in
the tagged potential energy power spectra of methanol, water and silica,
despite the differences in the underlying interactions and the dimensionality
of the network. All three liquids show a distinct multiple time scale (MTS)
regime with a $1/f^\alpha$ dependence with a clear positive correlation 
between the scaling exponent $\alpha$ and the diffusivity.
The low-frequency limit of the MTS regime is determined by the
frequency of crossover to white noise behaviour which occurs at approximately
0.1 cm$^{-1}$ in the case of methanol under standard temperature and
pressure conditions. The power spectral regime above 200 cm$^{-1}$ in all
three systems is dominated by resonances due to localised vibrations, such
as librations.  The correlation between $\alpha$ and the diffusivity in
all three liquids appears to be related to the strength of the coupling
between the localised motions and the larger length/time-scale network
reorganizations. Thus the time scales associated with
network reorganization dynamics appear to
	be qualitatively similar in these systems, despite the
	fact that water and silica both display diffusional
	anomalies but methanol does not.

\vfill

{* Author for correspondence (Tel: (+)~91~11~2659~1510; Fax: (+)~91~11~2686~2122; E-mail: {\tt charus@chemistry.iitd.ernet.in})\hfill}
\newpage

\section{Introduction}

Considerable experimental as well as theoretical attention has been 
devoted to characterizing the structure and dynamics of hydrogen-bonded
liquids \cite{ls93,bb03,bb05,mcbf99,lrt061,lrt062,rb05,rlp04,lc96,at07}.
The motivation for these studies is primarily due to the importance of water
as a solvent for biological and chemical processes. A comparison of the
behaviour of water with other second-row hydrides,
such as HF, NH$_3$ and CH$_3$OH, is essential for understanding
the nature of hydrogen-bonded dynamics and the extent to which the
behaviour of water is  unique. Methanol, in particular, is of
interest because it is the simplest molecule which can exhibit
both hydrogen bonding and nonpolar interactions.

While interactions in simple liquids are dominated by steep, short-range 
repulsions and long-range, isotropic attractions, hydrogen-bonded
liquids have strong, local anisotropic interactions. In the case of water, each
molecule can form at most four hydrogen bonds,  leading to a three-dimensional,
open, locally tetrahedral network structure. The strength of hydrogen
bonds is estimated to lie between 5 and 10$k_BT$ at melting \cite{jt99} which
is strong enough that a substantial fraction of hydrogen bonds will be intact 
at room temperature. 
Thermal fluctuations will, however,  be large enough, in comparison
to the bond strength,  to ensure that such bonds will have a finite lifetime 
of the order of picoseconds. As a result, the dynamics of the
liquid will be dominated by the
behaviour of the three-dimensional, hydrogen-bonded network, parts of which
are constantly broken and reformed.  In the case of methanol, hydrogen bond 
strengths
are similar to those in water \cite{wlj86}, but each methanol molecule
can form at most three hydrogen bonds, of which only one can be as a 
proton donor. In liquid methanol, however, simulations as well as
neutron scattering experiments show that the typical number of hydrogen
bonds per molecule is two \cite{yhs99,ybs00,wnfz00}. Consequently, linear 
chains with very rare branch points are seen, rather than the three-dimensional 
network characteristic of water. 

Hydrogen-bonded liquids can be 
thought of as a subset of network-forming liquids, other examples
being ionic melts such as ZnCl$_2$, SiO$_2$ and BeF$_2$ \cite{mw00}.
The strong local coupling of individual atomic or molecular units leads
to the existence of multiple length and time scales corresponding to
cooperative rearrangements  of the network involving different 
numbers of molecules.  To characterise the dynamical
behaviour of network-forming liquids, it is convenient to 
use the power spectral density of a mechanical quantity that is
sensitive to motion on a number of different length scales.
The power spectral density of an observable $A(t)$ as a function of time $t$
over a time interval $T$, is defined as\cite{wu45}
\be
\label{psd}
S(f) =\lim_{T\rightarrow\infty} \frac{1}{T} \left| \int_{-T/2}^{T/2} A(t)
e^{2{\pi}ift}dt\right|^2.
\ee
where $S(f)$ is the spectral power associated with frequency $f$. 
In a recent set of studies, we have shown that the power spectrum associated
with the tagged particle potential energy can be used to characterize the
dynamics of  tetrahedral, network-forming liquids, such as water 
\cite{mrc03,mc04,mc05,mc06,mcm06} and SiO$_2$ \cite{smc06}. 
Simple liquids show a
white noise power spectrum where S(f) is independent of frequency.
Network-forming liquids, however,  have a characteristic  multiple
time-scale regime with  $1/f^{\alpha}$ dependence on
frequency $f$ with $\alpha$ lying between 1 and 1.8 \cite{sor92}.
The three liquids that we have studied all show a region of
anomalous diffusivity where mobility increases on isothermal compression;
this density-dependent variation in mobility  is strongly correlated
with the scaling exponent $\alpha$. Our results therefore suggest that
monitoring the power spectrum of tagged particle potential fluctuations to
be a simple and direct method for linking phenomena on three distinctive 
length and time scales: the local molecular environment, hydrogen bond 
network reorganisations and the diffusivity.  

In this work, we characterise the multiple time-scale
behaviour of liquid methanol, focusing primarily on the 
power spectrum of the tagged particle potential energy.
Methanol presents an interesting contrast to the systems we have
studied earlier since hydrogen bonding results in formation of one-dimensional
chains, rather than a three-dimensional network. Unlike water,
methanol does not show any anomalous liquid state properties, such
as the diffusional anomaly mentioned above,  or thermodynamic anomalies
related to the existence of a temperature of maximum density under
isobaric conditions. It is therefore interesting to compare
the manner in which the multiple time-scale behaviour of methanol and
water differ. We perform molecular dynamics simulations of methanol
using Haughney's H1 potential \cite{hfm87} which predicts thermodynamic, structural and 
transport properties in fair agreement with a wide variety of experimental 
data\cite{yhs99,ybs00,wpgy96,an98,vcs00} as well as recent ab initio molecular 
dynamics simulations \cite{tkt99,hmg04}.
Section 2 contains  details of the molecular dynamics simulations as well
as of the power spectral analysis. 
Section 3 contains a discussion of our results, and Section 4 presents the 
conclusions.

\section{Computational Details}

\subsection{Potential energy surface}

The intermolecular  interactions in methanol are  modeled using 
Haughney's H1 pair-potential  which treats methanol as a rigid three-site entity
where the centres of interaction are positioned on the carbon (C), oxygen (O)
and the hydroxyl proton H$_O$\cite{hfm87}. The interaction between any two methanol molecules $\alpha$ and $\beta$ has the following functional form:
\be
U_{\alpha\beta}=\sum_{i}\sum_{j} 4\epsilon_{ij}\left[\left(\frac{\sigma_{ij}}{r_{ij}}\right)^{12}
-\left(\frac{\sigma_{ij}}{r_{ij}}\right)^6\right]
+ \frac{q_iq_j}{r_{ij}}
\ee
where the indices $i$ and $j$ label the atomic sites on the two molecules
and  $r_{ij}$ corresponds to the site-site separation. The atomic
sites are labelled by the partial charge, and the Lennard-Jones parameters
$\sigma_{ii}$ and $\epsilon_{ii}$, which are summarised in Table I. 
The Lennard-Jones parameters for interactions between unlike atoms are obtained 
using the Lorentz-Berthelot mixing rules:
$\sigma_{ij}=(\sigma_{i}+\sigma_{j})/2$ and
$\epsilon_{ij}=\sqrt{\epsilon_{i}\epsilon_{j}}$.
The molecular geometry is specified using the following bond lengths and
bond angles:  r(C-O)=1.4246${\mathring{A}}$, r(O-H$_O$)=
0.9451${\mathring{A}}$, r(C-H)=1.0936${\mathring{A}}$, $\angle$COH$_O$=
108.53$^{\circ}$ and $\angle$HCH =108.63$^{\circ}$.
The methyl protons are not considered as interaction sites but they are 
included when specifying the molecular geometry, so that the molecules have 
correct masses and moments of inertia. 

\subsection{Molecular Dynamics}

Molecular Dynamics (MD) simulations of a system of 256 methanol molecules
were carried out in canonical (NVT) ensemble, using the DL$\_$POLY software 
package \cite{sf96,syr01}, under cubic periodic boundary conditions. The 
effects of electrostatic,
 long-range interactions were accounted for by the Ewald summation method.
The non-coulombic part was truncated at half the box length. 
Berendsen thermostat, with a time constant $\tau_B$=200ps, 
was used to maintain the desired temperature for the production run.
 The Verlet algorithm with a time step of 2 fs was used to integrate the 
equations of motion. The quaternion algorithm was used to maintain the
rigid body constraints on the methanol molecules.

The system was studied along two isochores, 0.768 g cm$^{-3}$ and 
0.878 g cm$^{-3}$. The experimental density at 300 K and 
1atm is 0.768 g cm$^{-3}$ \cite{hfm87} while 0.878 g cm$^{-3}$ is the 
density around which the glass transition was reported to occur at around 
155 K using the H1 potential \cite{sk92}. 
The temperatures studied were 300, 250, 210, 180, 170 and 155 K  
for the 0.768 g cm$^{-3}$ isochore and 300, 250, 210, 180 and 170 K for
the 0.878 g cm$^{-3}$ isochore. 
The initial configuration for the molecular dynamics simulations at 300 K 
was generated using the PACKMOL program \cite{mm03}. The system was then slowly 
cooled to obtain lower temperatures. The system was equilibrated for 1-4 ns 
for the temperature range of 300-155K, and the production run lengths
were kept between 2-8 ns.
The details of run lengths are summarised in Table II.
Self-diffusivities were computed using the Einstein 
relation \cite{hm86}. Long run lengths were used, particularly at low temperatures, 
to ensure 
adequate sampling of configuration space, as measured by the mean squared 
displacement (MSD) of oxygen atoms, as shown in Figure 1. 
The time of onset of the diffusional regime, $t_{diff}$, was
identified from the MSD versus time plots as the regime in which the
Einstein relation is obeyed.  Along an isochore,
as the temperature is decreased we observe that the onset of diffusional regime 
shifts to higher and higher times (0.5 to 2.5 ns). At 155 K, for the 
0.768 g cm$^{-3}$ isochore, no linear dependence of the MSD on time is observed 
even upto 8 ns run length, suggesting that the system is out of equilibrium and 
we are dealing with an arrested structure.  Table II shows
the time of onset of diffusional regime, $t_{diff}$, as determined from
the applicability of the Einstein relation to describe the time
dependence of the MSDs.
Our simulation results are in agreement with previous results \cite{hfm87,sk92} 
within the statistical error bars.

\subsection{Generating Power Spectra}
 
Based on our previous
work on water where we considered various tagged particle observables, such
as local orientational order metrics and tagged particle potential
and  kinetic energies,
we concluded that the tagged particle potential energy is sensitive
to a wide range of time scales associated with the hydrogen-bonded
network \cite{mc04}. Therefore, in this study, we focus on the static 
distributions and power spectra associated 
with tagged molecule potential energies, $S_u(f)$.
The tagged molecule potential energy, $u(t)$, corresponds to the
interaction energy of an individual molecule with all the other molecules in the
system. Since the configurational potential energy function is pair-additive,
 the total potential
energy, $U(t) =0.5\sum_i u_i(t)$ where the sum extends over all molecules.

The tagged particle potential energies were stored at intervals of 10 fs
i.e. every 10 MD steps, which corresponds to a Nyquist frequency of 
1666 cm$^{-1}$. The value of
the Berendsen thermostat time constant ($\tau_B$) provides the lower limit on 
the frequency range over which we can obtain reliable power spectra; thus,
$\tau_B$=200 ps corresponds to a lower frequency limit of 0.165 cm$^{-1}$. 
Standard fast Fourier transform routines were use with a square sampling window.
The normalization convention  was chosen such that the integrated area under
the $S(f)$ curve equalled the mean square amplitude of the time signal. Windows
containing 2$^{16}$ data points were used for Fourier transformation
\cite{numrec}. 
Statistical noise in the power spectra was reduced by averaging overlapping
time signal windows as well as individual tagged molecule spectra \cite{mrc03}. 

Our  analysis of the power spectra associated with tagged molecule 
potential energy focuses on the three distinct regions of the spectrum: 
(i) identification of resonances or power spectral peaks, typically
corresponding to high-frequency local vibrational modes  which
are weakly coupled to network reorganizations; (ii) 
identifying the frequency range and the exponent associated
with the $1/f^{\alpha}$ multiple time scale regime, and
(iii) the frequency of crossover to white noise behaviour.
A linear least-square fitting of $\ln S(f)$  was used to obtain the scaling
exponent $\alpha$ associated with the multiple time-scale region. 

\subsection{Model-based fitting procedure for power-law-like spectra}

A  procedure for quantitatively fitting the
entire power spectrum, and not just the multiple time scale regime, to
a physically motivated parametric
model for the time-scale distribution associated with molecular motions in
water has been developed previously\cite{mcm06}. 
In this paper, we apply a similar procedure to quantify
the temperature-dependent changes that take place along the 0.768 and 0.878 
g cm$^{-3}$ isochore of H1 model of liquid methanol. The analytical model is a 
sum of six functions that correspond to well-defined model behaviors:
\begin{itemize}
\item  a $S_{\beta}(f;a, \beta) = a^2/f^{\alpha}$ power-law noise term  containing two parameters, corresponding to the  amplitude and spectral index;
\item five resonance distributions. Here we have taken a flat distribution of resonance frequencies lying between
$f_{min}$ and $f_{max}$ with fixed relaxation rate $\lambda$. 
The resulting spectral densities are
\begin{equation}
S_{res}(f; a,f_{min},f_{max},\lambda) =  a^2 \int_{f_{min}}^{f_{max}}
\left[ {\lambda^2 + 4\pi^2(f-f_0)^2} \right]^{-1} {df_0}.
\end{equation}
For each resonance, four  parameters, corresponding to the
amplitude, minimum and maximum resonance frequencies and relaxation rate, must
be introduced.
\end{itemize}
In addition the model function also contains a white noise term and  a filter 
function 
\begin{equation}
F(f; \lambda,n) = \frac{1}{((2\pi f/\lambda)^{2n} + 1)}
\end{equation}
with $n \approx 5$. 
The filter function takes into account the steep fall off in the spectral
density at high frequency, presumably due to the upper limit imposed
by the time step of the molecular dynamics algorithm. The filter function 
is approximated rather roughly by the function $F(f; \lambda,5)$, the numerical tests show that that the fit is rather insensitive to its exact shape. 
The overall 24-parameter model is given by:
\begin{equation}
\nonumber
S_{model}(f) = F(f; \lambda_0,5) \left[ S_{\beta}(f;a_1, \beta_1) + \sum_{k=2}^{k=6} S_{res}(f; a_k,f_{min,k},f_{max,k},\lambda_k) \right] + C
\end{equation}
where $C$ is an additional white noise term.

\section{Results and Discussion}

\subsection{Static distribution of tagged molecule potential energy}

Figure 2 shows the static distribution, $P(u)$, of the tagged molecule
potential energies, $u(t)$, for  methanol at several temperatures along the 
0.768 and 0.878 g cm$^{-3}$ isochores. Unlike in the case of monoatomic
van der Waals liquids, methanol resembles water and silica
in that, it shows multimododal $P(u)$ distributions indicating that
there are energetically distinct local environments in the liquid.
The presence of distinct peaks suggests that there is a small energetic 
barrier to transit from one energetic environment to another. This 
heterogeneity in local environments seems to be characteristic
of network-forming liquids, regardless of the dimensionality of the
networks. It should be noted that the number of
peaks in the $P(u)$ distributions of water and silica are larger but
the individual peaks are less distinct.

The three peaks in methanol are centred at approximately at $\sim$ -130, -85 
and -30 kJ mol$^{-1}$. The distinct demarcation between the
three peaks as well as the relative probabilities 
change with temperature, with the lowest energy peak becoming 
less prominent. At low temperatures, just above the
glass transition, the relative intensities of the peaks are very sensitive to
the quality of equilibriation. The mean value of the tagged potential
energy, $\expt{u}$, varies between -82.4 and -69.2 kJ mol$^{-1}$ between 155 
and 300 K along the 0.768 g cm$^{-3}$ isochore.
A similar behaviour is observed in case of 0.878 g cm$^{-3}$ isochore. The mean 
value of the tagged potential energy, $\expt{u}$, ranges from 
-84.17 to -72.3 kJ mol$^{-1}$ between 170 and 300 K along this isochore.

Figure 3 shows the probability distributions corresponding to the
tagged molecule potential energy contributions from the
van der Waals, screened Coulombic and reciprocal space contributions
to the total potential energy for a single state point (300K, 0.768 g cm$^{-3}$).
It is immediately obvious that
the multimodal character is due to the electrostatic contribution
from the screened Coulombic interactions. The
$P(u_{rec})$ and $P(u_{vdW})$ distributions are unimodal and
very narrow  in comparison with $P(u_{screen})$.

The dependence of the multimodality of the static tagged particle potential 
energy distribution on the electrostatic contribution was also noted in our 
previous work on silica \cite{smc06}. In the Haughney model, the Coulombic 
contribution is parameterised to take into account the net effect of both 
hydrogen bonding and multipole moment interactions. To understand the extent 
to which the
peak structure of the $P(u)$ distributions is correlated with the hydrogen
bonding, we have classified the methanol molecules based on their
tagged particle potential energy into three categories, as defined in Table III,
and computed the fraction $f_n$ of $n$-bonded hydrogen molecules in each 
category. 
The geometric criteria for identifying the hydrogen bonds were taken from
ref.\cite{hfm87}. Two molecules are regarded to be 
hydrogen bonded if $r_{OO}$ $\le$ 3.5${\mathring{A}}$, $r_{OH_O}$ $\le$ 
2.6${\mathring{A}}$ and the angle $\angle$$H_O$OO $\le$ 30$^{\circ}$.
The simulation averages of $f_0$, $f_1$, $f_2$ and $f_n$ at 300K and 0.768 g,
shown in Table III, reproduce those of Haughney et. al. within statistical
error. 
 Activation energes of hydrogen bond lifetimes was found to be 15.6 
kJ mol$^{-1}$, which is close to the previously reported values
of ~16 kJ mol$^{-1}$ \cite{hfm87,sk92}.

The results in Table III show that non-bonded molecules are essentially
absent in the 170K to 300K regime. 
The peak centered at -30 kJ mol$^{-1}$ has a somewhat
higher value of $f_1$ and the peak
centered at -130 kJ mol$^{-1}$ has a somewhat higher value of $f_3$ than the
overall simulation averaged values of $f_1$ and $f_3$.
The CH$_3$OH molecules which belong to the central peak of the $P(u)$
distribution have an $f_n$ distribution which is essentially identical
to the overall ensemble averaged distribution. Therefore the peak structure of
the $P(u)$ distributions cannot be attributed to differing degrees of
hydrogen bonding, though there is a weak correlation between $n_{hb}$ and
$u$. Our previous work on water, as well as the results in Section 3.4,
show that as the electrostatic contribution is turned off and the
Lennard-Jones component relatively enhanced, the degree of multimodality
diminishes. The multimodality must therefore originate from longer range
electrostatic interactions, stretching over at least next nearest neighbours,
in addition to nearest neigbours. It would be  interesting, in this context,
to study fluids with dipolar, and perhaps higher order multipolar,
interactions \cite{hm86}.

\subsection{Power spectra of tagged particle potential energy}

\subsubsection{Comparison of methanol and water}

Figure 4 compares the power spectra generated by the fluctuations in the
tagged potential energy of methanol, $S_u(f)$, at 300 K and 0.768 g cm$^{-3}$
with that of SPC/E water at 1.0 g cm$^{-3}$ at 300 K. The densities correspond
to those observed under ambient temperature and pressure conditions.
Despite the difference in the dimensionality of the hydrogen bonded network
between methanol and water, the key power spectral features show strikingly
similar qualitative features with: 
(i) high frequency localised modes 
(ii) a $1/f^\alpha$ multiple time scale regime with exponent
$\alpha$ at intermediate frequencies and (iii) a crossover to white noise
at low frequencies. The high-frequency cutoff of the power spectrum is
determined by the time step used in the molecular dynamics simulations.

Given the rigid molecule approximation used in the present simulations,
no intramolecular vibrational modes can be observed in our studies. 
We briefly summarise
the information on specific intermolecular vibrational modes available
from previous experimental and simulation studies.  In the case of water, 
there is a broad librational band between 500 and 800 cm$^{-1}$,
as well as two-molecule O-O stretching and three-molecule O-O-O bends
at approximately 200 cm$^{-1}$ and 60 cm$^{-1}$ \cite{mi86,bg91,ph98}.
In the case of methanol, experimental IR spectra locate peaks at approximately
680, 270, 135 and 65 cm$^{-1}$ \cite{vcs00,pkm70,gmo90}. The hindered
rotations of the methanol molecule, with the methyl group treated
as a united atom in the OPLS model \cite{wlj86}, are associated with frequencies
of approximately 580, 210 and 50 cm$^{-1}$ \cite{mg90}.  
Explicit treatment of the geometry of the methyl group in the 
H1 potential shows that  the 270 cm$^{-1}$ peak is associated with
the velocity autocorrelation function of the hydroxyl hydrogen atom
\cite{hfm87}. More recent studies of the optical spectra of methanol
identify the COH pure librational peak as lying between 200 and 300 cm$^{-1}$
and the peak at approximately 600 cm$^{-1}$ as the  H-bond constrained
libration around the CO bond \cite{cccrp99,pcrs03}. The low frequency
librational bands at 65 and 135 cm$^{-1}$ are expected to be strongly
coupled with intermolecular vibrational modes.

The $S_u(f)$ spectrum of water shows a broad librational band between 
500 and 800 cm$^{-1}$ but the intermolecular stretches are not
observable at this density and temperature since they lie in the
multiple time-scale (MTS) region. In the case of methanol, the two high 
frequency librational bands are observed as distinct peaks at 270 and 
600 cm$^{-1}$.  Lower frequency vibrational bands at 65 and 135 cm$^{-1}$
are not visible as distinct peaks in the $S_u(f)$ spectrum of CH$_3$OH at this 
state point because they lie in the MTS region.

The multiple-time scale (MTS) regime in methanol at 300 K and 0.768 g 
cm$^{-3}$  extends from 0.2 to 200 cm$^{-1}$. There is, however, a
small but distinct change in the  exponent $\alpha$ at 40 cm$^{-1}$;
the $\alpha$ value for 0.2-40 cm$^{-1}$ range is 1.36 while for 40-200 
cm$^{-1}$ it is 1.22.
 For SPC/E water at 300 K and 1.0 g cm$^{-3}$, the $1/f^\alpha$ region extends 
from 1-200 cm$^{-1}$ with $\alpha$ =  1.4.
The presence of a larger number of high-frequency resonances in methanol,
compared to water, maybe responsible for this change in the exponent of
the MTS regime at approximately 40 cm$^{-1}$. In the remainder of this
paper, we refer as the MTS regime to the frequency interval
lying below 40 cm$^{_1}$ and above the frequency of crossover to white noise. 

At the equilibrium densities associated with
1 atm pressure and 300K, the white noise regime in water 
occurs below 1 cm$^{-1}$ whereas for methanol, it begins at a lower 
frequency of 0.2 cm$^{-1}$. This suggests that the time taken for the potential
energy fluctuations to decorrelate is smaller for water than methanol
and therefore the overall relaxation dynamics over length scales of a 
few molecules will be faster in water than in methanol. This is
consistent with the viewpoint expressed by Ladanyi and Skaf \cite{ls93}, based
 on orientational correlation functions, angular VACFs and  NMR
relaxation rates, that water and methanol relax at similar scales at short times
but the relaxation of CH$_3$OH becomes much slower for times greater than
0.6ps (55.5 cm$^{-1}$). The overall faster dynamics of water, despite the 
existence of a stronger hydrogen bond and an extended three-dimensional
network is attributed to the higher translational diffusity
due to lighter mass, more rapid librational modes and the  coincidence
of the direction of the molecular dipole moment with one of the principal
axes of rotation \cite{ls93}.

The power spectra associated with the fluctuations in the total tagged potential
 energy and the contributions from reciprocal and screened energies arising out 
of Ewald summation and the van der Waals interactions were computed
at 300K and 0.768 g cm$^{-3}$. 
Though the figure is not shown, the behaviour of $S_u(f)$ is almost exactly 
reproduced over much of the frequency range by the screened charge contribution.
 The spectral power associated with the reciprocal space and van der Waals 
contributions is lower and, therefore has a minor effect on the overall spectrum
 except in the high-frequency region. The $u_{vdW}$ contribution produces a 
relatively unstructured power spectrum, more like Argon, while the Coulombic 
interactions are responsible for much of the high-frequency, short-length scale 
structure. This is consistent with our earlier observations on silica \cite{smc06}.

\subsubsection{Temperature and Density Dependence}

The temperature dependence of the power spectra associated with tagged 
molecule potential energies is considered in Figure 5. Figures 5(a)  and
5(b) show $S_u(f)$ curves for different temperatures along the 0.768 and
0.878 g cm$^{-3}$  isochores respectively. The glass transition, as reflected
in the time-dependence of the mean-square displacement plots, occurs 
at a lower temperature for the higher density isochore.

The $S_u(f)$ plots for both densities at 300 K show well-defined librational 
peaks. As discussed in section 3.2.1, the  $1/f^\alpha$  MTS region
can be partitioned into a low-frequency MTS regime 
extending from 0.2-40 cm$^{-1}$ with $\alpha_u$ =  1.36
and a higher frequency MTS regime from 40-200 cm$^{-1}$ with exponent 1.22.

At this temperature, there are no peaks lying between 40 and 200 cm$^{-1}$
though molecular centre-of-mass VACFs, far-IR and dielectric loss spectra 
suggest that vibrational
modes at 65 and 135 cm$^{-1}$ corresponding to intermolecular vibrational
modes involving two or three molecules should be present.
The change in the value of $\alpha$ at approximately 40 cm$^{-1}$ suggests that 
these two- or three- molecule network modes are  partially decoupled from the 
longer length scale, lower frequency network reaarrangements. 
This discontinuity becomes more pronounced with the lowering of temperature and 
distinct peaks at approximately 65 and 140 cm$^{-1}$ can be seen in $S_u(f)$ 
plots at 170 and 155K. These peaks must correspond to caging vibrations of 
methanol molecules as the system dynamics slows down.  For these low temperature
 isotherms, the onset of diffusional dynamics in the MSD plots is very slow 
indicating that the system is very close to glass transition.
The crossover to white noise which is visible below 0.2 cm$^{-1}$ for the 300K 
isotherm moves to  frequencies  below 0.1 cm$^{-1}$ for temperatures of
250K or less. In this study, we did not attempt to generate power spectra
at lower frequency since this would have implied using very large values
for the time constant of the Berendsen thermostat and therefore would have
compromised the temperature control.

\subsection{Parametric fitting of the Power Spectra}

The results of the fitting of the power spectra along the 0.768 g cm$^{-3}$ 
isochore is shown in Figure 6. The quality of the fits, given the complex
shape of the power spectrum, is good. It is interesting to compare the
parametric distribution functions used for water \cite{mcm06} and methanol.
Fitting the water power spectra requires only two resonances \cite{mcm06}
while fitting the methanol power spectrum requires five resonance
terms. In the case of both water and methanol, power spectra were
computed over the 0.1 to 1666 cm$^{-1}$ regime.
The low frequency white noise  is more pronounced for water in 
this frequency regime than for methanol for the temperatures and 
densities studied here. As a consequence, a Debye term corresponding to 
fixed frequency relaxation processes is required for the
 water spectral fits.
In the case of methanol,  an additional white noise term and a 
filter function is used in the fits but no Debye relation term was required.

\subsection{Modifying the Strength of Hydrogen Bonding}

The H1-model of methanol effectively condenses the information about the
short-range repulsion and dispersion using the Lennard-Jones terms. The long-
range electrostatic interactions and anisotropic hydrogen bonding interactions
are controlled by the partial charge distribution over the atoms. To study the
 effect of varying the strength of hydrogen bonding on the power spectra and the
 static distribution of tagged particle potential energies at 300 K and 0.768
g cm$^{-3}$ , we have scaled the H1-model charges.
Figure 7 compares the $S_u(f)$ spectra of methanol with
normal charges on C, H$_O$ and O with the power spectra when the charges are
up-scaled and down-scaled by 10\%, at 300 K and 0.768 g cm$^{-3}$. The crossover
 to white-noise occurs at around 0.2 cm$^{-1}$ for the system with standard
 H1 model charges. When  the charges are increased  by 10\%,
 we see that the librational peak becomes
more pronounced and the crossover to white noise shifts to
lower frequencies (not shown in the frequecy range studied) owing to the 
fact that now the
interactions are much stronger and so is the hydrogen bonded network resulting
in a long range correlation between the network dynamics and the local
librations in the system. An increase in the extent of multimodality in the
probability distribution of static tagged potential energy is also observed with
 the increase in charges on H1 model of methanol. 
On the other hand, reduction of the charges by 10\%, converts a sharp
librational peak occuring in tagged potential energy power spectrum to almost
a shoulder and induces a crossover to white noise at frequencies below
1 cm$^{-1}$. There also occurs a destructuring or loss of multimodality in
$P(u)$ consistent with the weakening of the hydrogen-bonded network in liquid
methanol.
 As the charges are reduced and the system approaches a Lennard-Jones fluid in
its behaviour, the tagged molecule potential energy distribution narrows
significantly indicating that the heterogeneity in local molecular
environments responsible for multiple time-scale behaviour is lost.

\subsection{Connection between multiple time scale behaviour and diffusivity}

The self-diffusivity was computed using the Einstein relation from the
MSD plots for the oxygen atoms (see Section 2.2). Figure 8 shows the
Arrhenius plot of the diffusivity. The activation energies along the
0.768 and 0.878 g cm$^{-3}$ isochores are 10.5 and 14.5 kJ mol$^{-1}$ respectively. 
Haughney et al report an activation energy of 13 kJ mol$^{-1}$ for 0.768 g cm$^{-3}$ isochore; the discrepancy
with our values may be due to the fact that our run lengths are more than
two orders of magnitude longer.

In simple liquids, diffusion occurs as a combination
of collisional and cage relaxation processes. An additional mechanism of
diffusion exists in network-forming liquids due to the coupling of
local high-frequency modes (e.g. librations in the case of H$_2$O) with
local, two-molecule  network re-organizations. As librational modes are more 
strongly coupled with network-reorganizations, the exponent $\alpha$ rises
as does the diffusivity, leading to a strong positive correlation
between $D$ and $\alpha_u$, as discussed in our earlier work on water and
silica \cite{mc05,smc06}. 
Figure 9 demonstrates that this relationship between $\alpha_u$ and
the diffusivity also holds in methanol. Clearly, the librational coupling to
larger scale H-bonded network reorganisations is an important diffusional
mechanism in CH$_3$OH, as in water and silica, despite the differences in
network dimensionality. It should be noted that we have computed $\alpha_u$
values by fitting $\ln S_u(f)$ over the 1 to 40 cm$^{-1}$ regime which 
is expected to be particularly sensitive to the coupling of the 
two- and three-molecule modes with the network regorganisations.

In order to make a quantitative comparison of the correlation between
diffusivity and the scaling exponent, Figure 10 plots a dimensionless
diffusivity, $D^*$, against $\alpha_u$ for  CH$_3$OH, H$_2$O and SiO$_2$
in the liquid phase. The dimensionless diffusivity, $D^*$, is defined as
$D^*={D \rho^{1/3}}/{\sqrt {k_BT/m}}$ where $\rho$ and $m$ correspond to
the number density and molecular mass of the 
formula units (H$_2$O, SiO$_2$ and CH$_3$OH).
The correlation is very similar in all three cases, though given the
differences in intermolecular interactions, some scatter in the correlation
plot is expected. Typically, as temperatures and densities increase and
the network structure breaks down, the $\alpha$ values  become 
less sensitive to the diffusivity; in the limiting case of a simple liquid,
$\alpha_u$ is zero and uncorrelated with the diffusivity.

\section{Conclusions}

In this work, we have focused on understanding the structure and dynamics
of liquid methanol by studying the static distributions and temporal 
correlations associated with fluctuations in the tagged molecule potential
energies. While the tagged molecule potential energy is not a directly
accessible quantity, previous work on water, aqueous alkali halide solutions
and silica melt have demonstrated that this quantity is sensitive to network
reoganisations on several time scales.  In the case of methanol, we show that 
analysis of this single, frequency-dependent power spectral 
distribution  provides an overall 
view of the dynamics, including important resonances and the multiple 
time-scale regime associated with network formation. 
This is particularly interesting given that the resonances at 650 cm$^{-1}$, 
270 cm$^{-1}$, 135 cm$^{-1}$ and 65$^{-1}$ have otherwise been identified
using very different correlation functions corresponding
to different experimental techniques. 

The static distribution of tagged molecule potential energies shows a
clear multimodal structure with three distinct peaks. Similar multimodal
distributions were observed earlier in the case of water and silica melt
at low temperatures and densities. For both the systems, factors which
attenuate the network, such as temperature, density and or ionic solutes,
were found to destroy the multimodal character and lead to unimodal
Gaussian-type distributions characteristic of simple liquids.
In the case of methanol, we have further found that the correlation 
between number of hydrogen bonds, as identified by geometric criteria, 
and the tagged particle potential energy is very weak. The multimodality
seems to therefore be due to electrostatic effects, but not necessarily
due to local anisotropic interactions. The longer range of electrostatic
interactions, such as dipole-dipole interactions, presumably implies
that the organisation of second and third neighbour shells plays
a crucial role in determining the tagged particle potential energy
distributions. In this context, it would therefore
be interesting to examine strongly dipolar, but not necessarily 
hydrogen-bonded systems.

An interesting outcome of this study is the remarkable similarity in
the tagged potential energy power spectra of methanol, water and silica,
despite the differences in the underlying interactions and the dimensionality
of the network. All three liquids show a distinct multiple time scale (MTS)
regime with a $1/f^\alpha$ dependence with a clear positive correlation 
between the scaling exponent $\alpha$ and the diffusivity.
The low-frequency limit of the MTS regime is determined by the
frequency of crossover to white noise behaviour which occurs at approximately
0.1 cm$^{-1}$ in the case of methanol under standard temperature and
pressure conditions. The power spectral regime above 200 cm$^{-1}$ in all
three systems is dominated by resonances due to localised vibrations, such
as librations.  The correlation between $\alpha$ and the diffusivity in
all three liquids appears to be related to the strength of the coupling
between the localised motions and the larger length/time-scale network
re-organizations. Thus the time scales associated with
network reorganization dynamics do not appear to
	be qualitatively different in these systems, despite the
	fact that water and silica both display diffusional
	anomalies but methanol does not. This suggests that the
	equilibrium transport properties  associated
	with the anomalies may be more sensitive to
		the way a liquids samples the overall configuration space,
		i.e. the excess entropy scaling of transport properties,
		than the  short time dynamics. Given that time-resolved
		infra-red techniques have been used to monitor the 
		hydrogen-bond dynamics in water, it would be interesting
		to compare the solvation dynamics in water with that
		in methanol and other low-molecular weight alcohols.

{\bf Acknowledgements} CC would like to thank the Department of Science and
Technology, New Delhi, for support under the Swarnajayanti Fellowship scheme. 
RS thanks Council of Scientific and Industrial research for the award of Senior
Research Fellowship.

\newpage
{\bf Table I}

{Potential Parameters of Haughney's H1 model}\\

\bigskip
\begin{tabular}{cccc}
\hline\hline
Atom     & $\epsilon$(kJ/mol)     & $\sigma$($\mathring{A}$)     &  $q$(e)  \\
\hline
C        &   0.7576               &   3.861                      &  0.297  \\
O        &   0.7309               &   3.083                      & -0.728  \\
$H_O$    &                        &                              &  0.431  \\
\hline\hline
\end{tabular}

\newpage
{\bf Table II}
                                                                                
{Summary of Computational Details for MD simulations of H1 methanol.
The lengths of equilibration and production runs are denoted by
$t_{eq}$ and $t_{prod}$ respectively. The time of onset of the
diffusional regime is denoted by $t_{diff}$.}\\
                                                                                
\bigskip
\begin{tabular}{c|ccc|ccc}
\hline\hline
Temperature  &              &0.768 g cm$^{-3}$ &                &              & 0.878 g cm$^{-3}$&               \\
(K)          & $t_{eq}$(ns) & $t_{prod}$ (ns) & $t_{diff}$ (ns) & $t_{eq}$(ns) & $t_{prod}$ (ns) & $t_{diff}$ (ns) \\
\hline
300          & 1            &  2              &      0.05       &  1           & 2           &  0.05              \\
250          & 1            &  2              &      0.08       &  1           & 2           &  0.10              \\
210          & 2            &  4              &      0.20       &  2           & 4           & 0.40              \\
180          & 3            &  6              &      2.00       &  3           & 6           & 3.00              \\
170          & 3            &  6              &      2.50       &  3           & 6           & 3.00              \\
\hline\hline
\end{tabular}
             
\newpage
{\bf Table III}
                                                                                
{Hydrogen bond statistics of methanol molecules at two different state points.
The fraction of molecules with $n$ hydrogen-bonded neighbours is
denoted by $f_n$, The ensemble averaged values of $f_n$ are coomputed, in
addition to the averages over molecules belong to restricted ranges
in the tagged molecule potential energy, $u$.}\\
                                                                                
\bigskip
\begin{tabular}{cccccc}
\hline\hline
                  &  300 K, 0.768 g cm$^{-3}$ &     &      &     &  \\
$u$ (kJ mol$^{-1}$) & $\expt{n_{hb}}$ & $f_0$ & $f_1$ & $f_2$ & $f_3$ \\
\hline
$u$ $\ge$ -48 & 1.68 & 0 & 0.28 & 0.63 & 0.05  \\
-48 $<$ u $\ge$ -103 & 1.87 & 0 & 0.19 & 0.72 & 0.08  \\
$u$ $<$ -103 & 2.12 & 0 & 0.04 & 0.80 & 0.16  \\
ensemble average & 1.86 & 0.017 & 0.19 & 0.71 & 0.082   \\
\hline
 & & & & & \\
\hline
                  &  170 K, 0.768 g cm$^{-3}$ &     &      &     &  \\
$u$ (kJ mol$^{-1}$) & $\expt{n_{hb}}$ & $f_0$ & $f_1$ & $f_2$ & $f_3$ \\
\hline
$u$ $\ge$ -54 & 1.91 & 0 & 0.105 & 0.88 & 0 \\
-54 $<$ u $\ge$ -108 & 1.98 & 0 & 0.05 & 0.92 & 0.028 \\
$u$ $<$ -108 & 2.07 & 0 & 0.0056 & 0.92 & 0.07 \\
ensemble average &   1.99 &0.00013 & 0.04 & 0.92 & 0.037 \\
\hline\hline
\end{tabular}
                                                               
                                                                                
                                                                                

\newpage
\begin{center}
{\bf Figure Captions}
\end{center}

\begin{enumerate}

\item Mean square displacements (MSDs) for different temperatures along (a) 
0.768 g cm$^{-3}$ and (b) 0.878  g cm$^{-3}$ isochores. Note the logarithmic 
scale along both the axes of the plot.

\item Static distribution, $P(u)$, of tagged potential energy, $u$, of 
	methanol, along (a) 0.768 g cm$^{-3}$ and (b) 0.878  g cm$^{-3}$ 
	isochores at different temperatures.

\item Contributions of reciprocal space, screened Coulomb and 
van der Waals interactions to the static distribution of tagged particle 
potential energy of methanol at 300 K and 0.768 g cm$^{-3}$. $P(u_{rec})$ has 
been scaled by 1/12 while $P(u_{vdw})$ has been scaled by a factor of half.

\item Comparison of power spectra associated with temporal fluctuations in
tagged particle potential energy, S$_u$(f), between methanol (H1 potential) and 
water (SPC/E potential) at 300K. The densities of methanol and water were
taken as 0.768 g cm$^{-3}$ and 1 g cm$^{-3}$, corresponding to experimental
densities at 1 atmosphere pressure.

\item Power spectra of tagged particle potential energy fluctuations, $S_u(f)$,
 of methanol, at different temperatures along (a) 0.768 g cm$^{-3}$ and (b)
0.878 g cm$^{-3}$ isochores. Relative magnitudes of power spectra for different
temperatures were scaled for clarity of presentation.

\item Power spectra of tagged potential energy, $S_u(f)$ of methanol
	fitted using the parameteric model defined in Section 2.3
along the 0.768 g cm$^{-3}$ isochore at (a) 300K and (b) 155 K.
The raw power spectral data are shown with points and the overall fit is shown
in solid lines.

\item Effect of modifying the partial charge distribution of the H1 model 
	potential on: (a) Power spectra of fluctuations of
tagged particle potential energies, $S_u(f)$; (b) Static distributions of tagged
particle potential energies at 300 K and 0.768 g cm$^{-3}$. Relative 
magnitudes of power spectra, $S_u(f)$, are scaled for clarity.

\item Arrhenius plot of self diffusion coefficient of oxygen along 0.768 and
0.878 g cm$^{-3}$ isochores.

\item Correlation plot between the self-diffusion coefficient of oxygen  
and  the scaling exponent,
$\alpha_u$ of the multiple time-scale  region of the $S_u(f)$ power
spectra in methanol. Diffusivities are in units of cm$^2$ sec$^{-1}$.

\item Correlation plot between the self-diffusion coefficient of oxygen in 
methanol and water, and silicon ion in silica with the scaling exponent,
$\alpha_u$ of the multiple time-scale  region of the $S_u(f)$ power
spectra. Dimensionless  diffusivities,  
$D^*={D \rho^{1/3}}/{\sqrt {k_BT/m}}$ where $\rho$ and $m$ correspond to
the number density and masses of the formula units have been used to
facilitate comparison between the different systems.

\end{enumerate}

\newpage

\begin{figure}
\caption{\quad}
\includegraphics[trim=1.5truein 0 0 2truein]{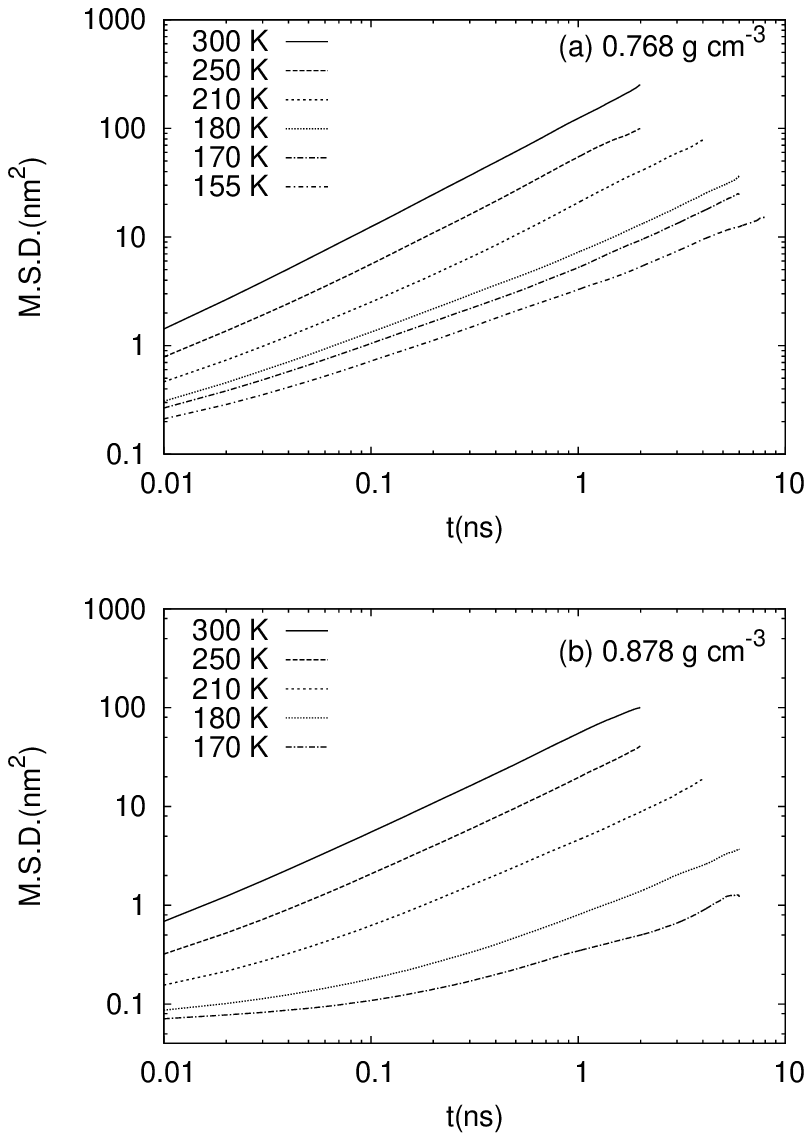}
\end{figure}

\newpage
\begin{figure}
\caption{\quad}
\includegraphics[trim=1.5truein 0 0 2truein]{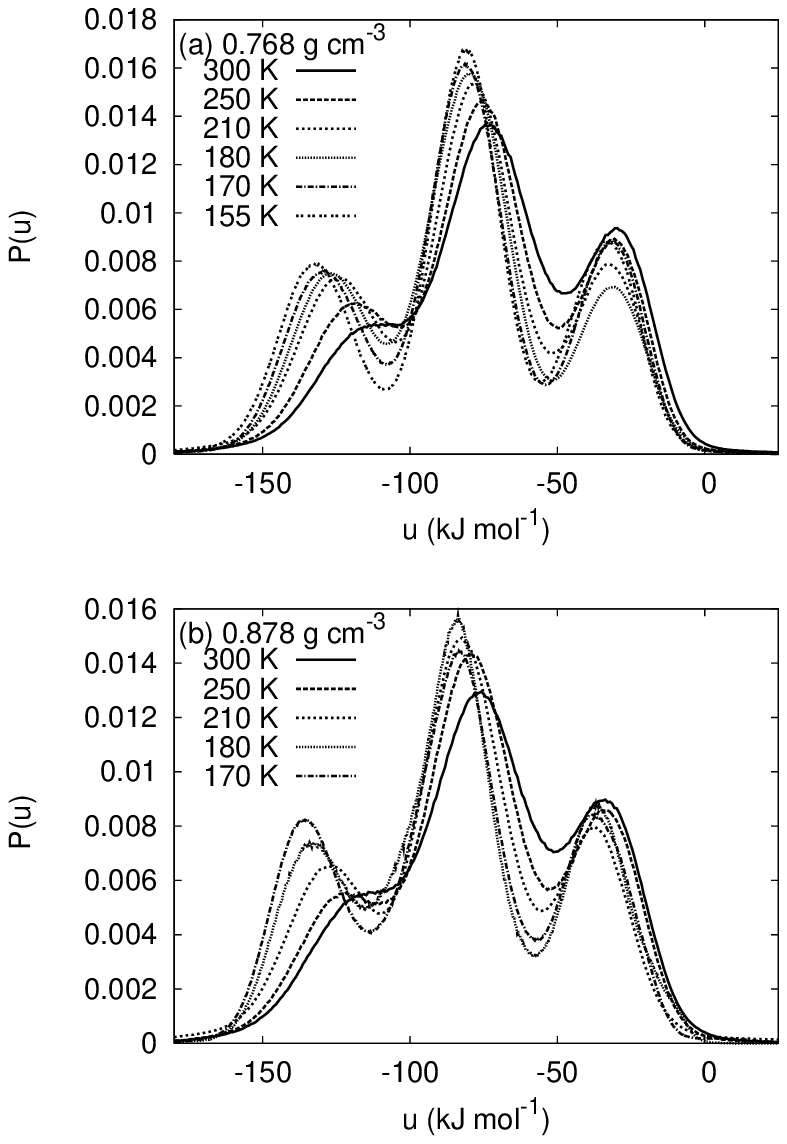}
\end{figure}

\newpage
\begin{figure}
\caption{\quad}
\includegraphics[trim=1.5truein 0 0 2truein]{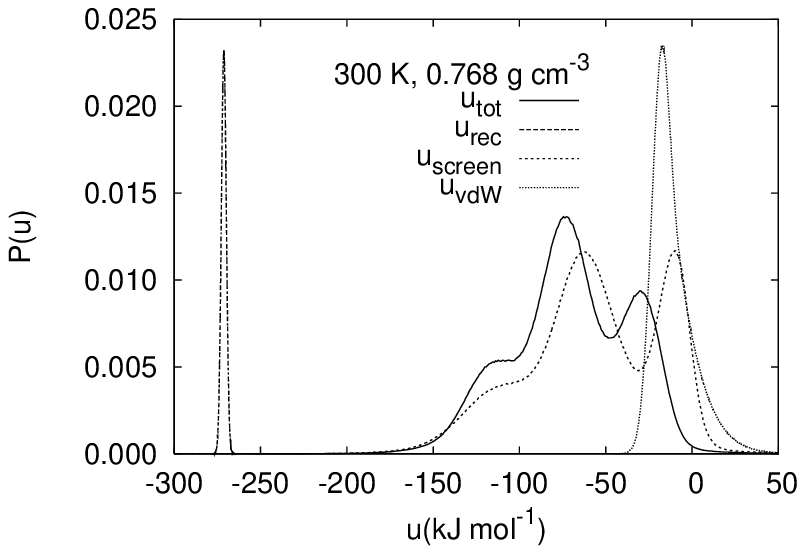}
\end{figure}

\newpage
\begin{figure}
\caption{\quad}
\includegraphics[trim=1.5truein 0 0 2truein]{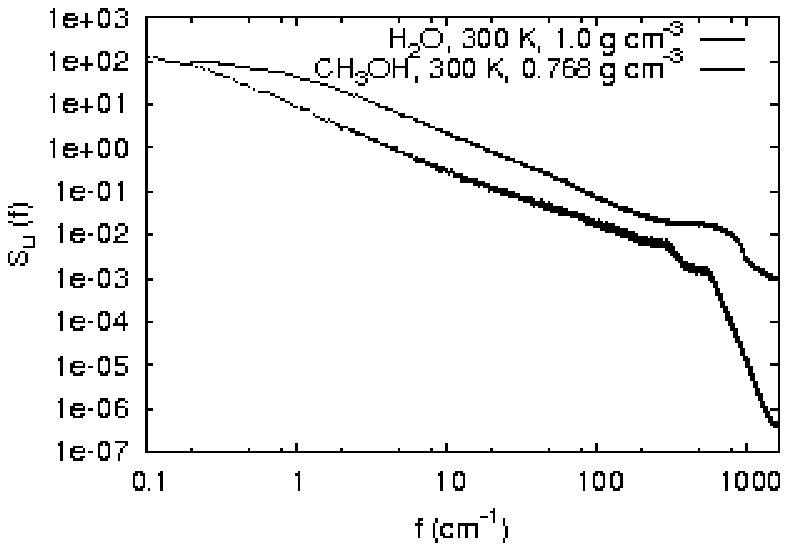}
\end{figure}

\newpage
\begin{figure}
\caption{\quad}
\includegraphics[trim=1.5truein 0 0 2truein]{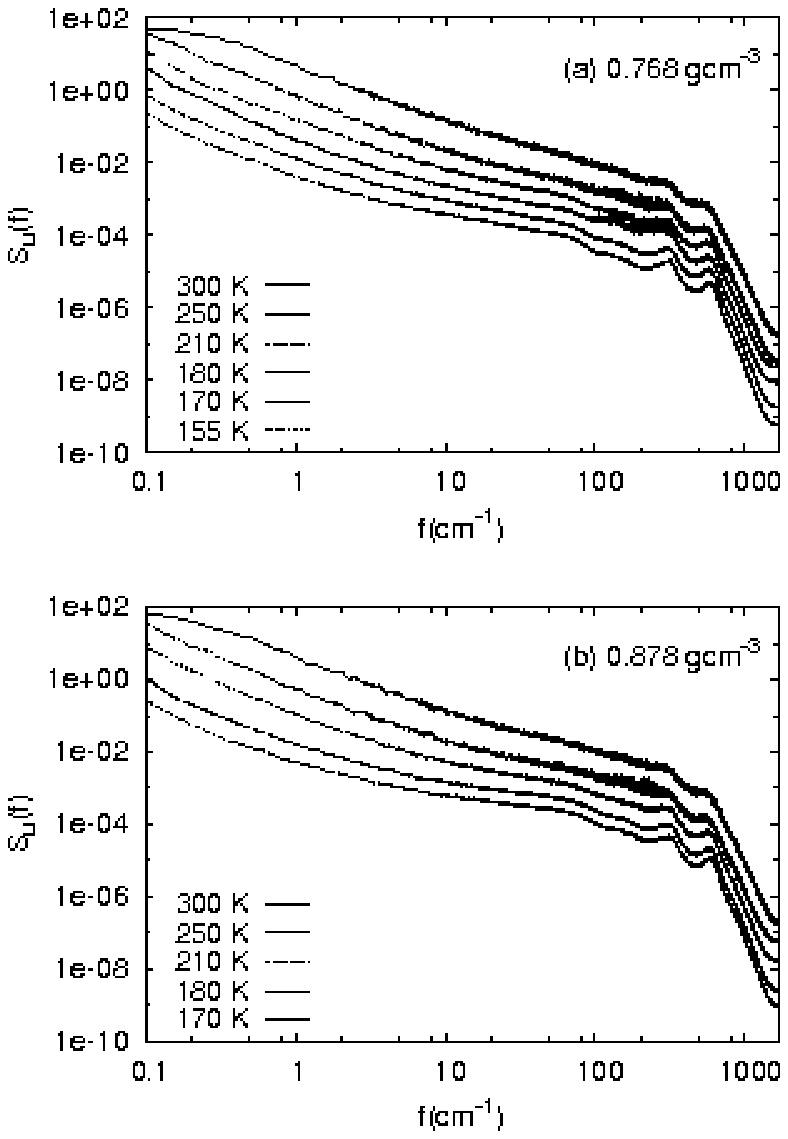}
\end{figure}

\newpage
\begin{figure}
\caption{\quad}
\includegraphics{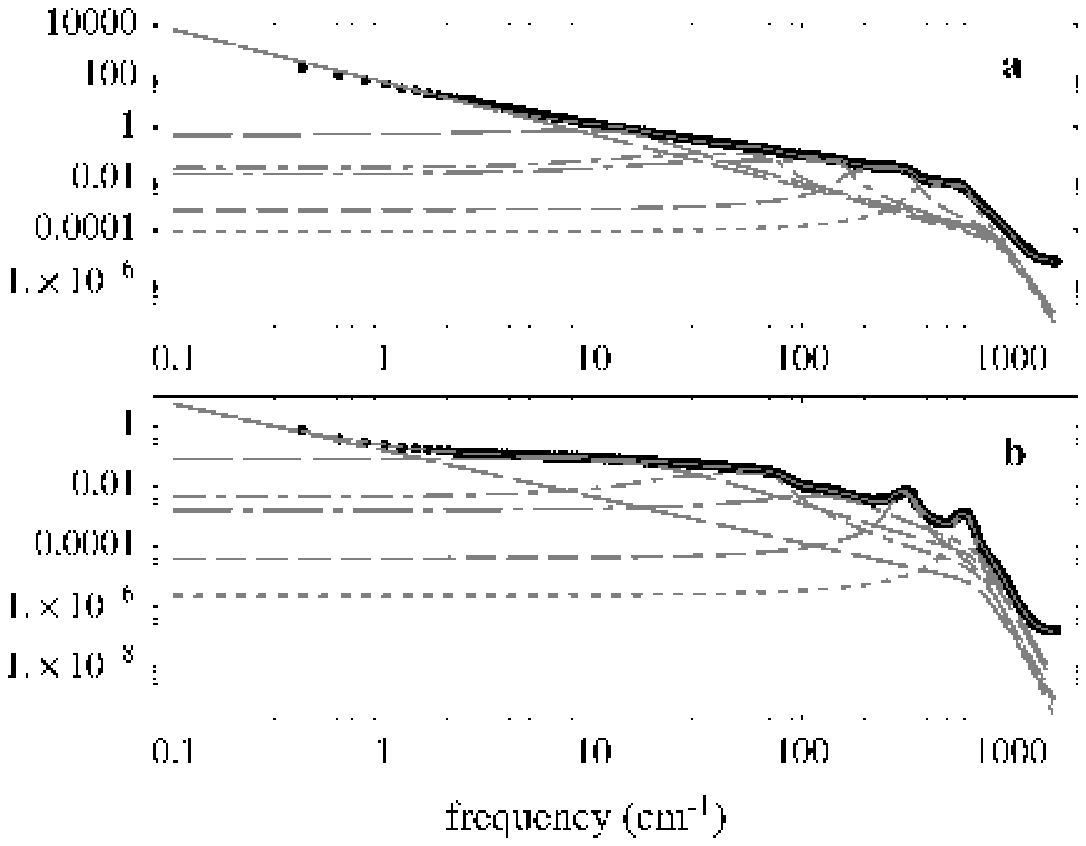}
\end{figure}

\newpage
\begin{figure}
\caption{\quad}
\includegraphics[trim=1.5truein 0 0 2truein]{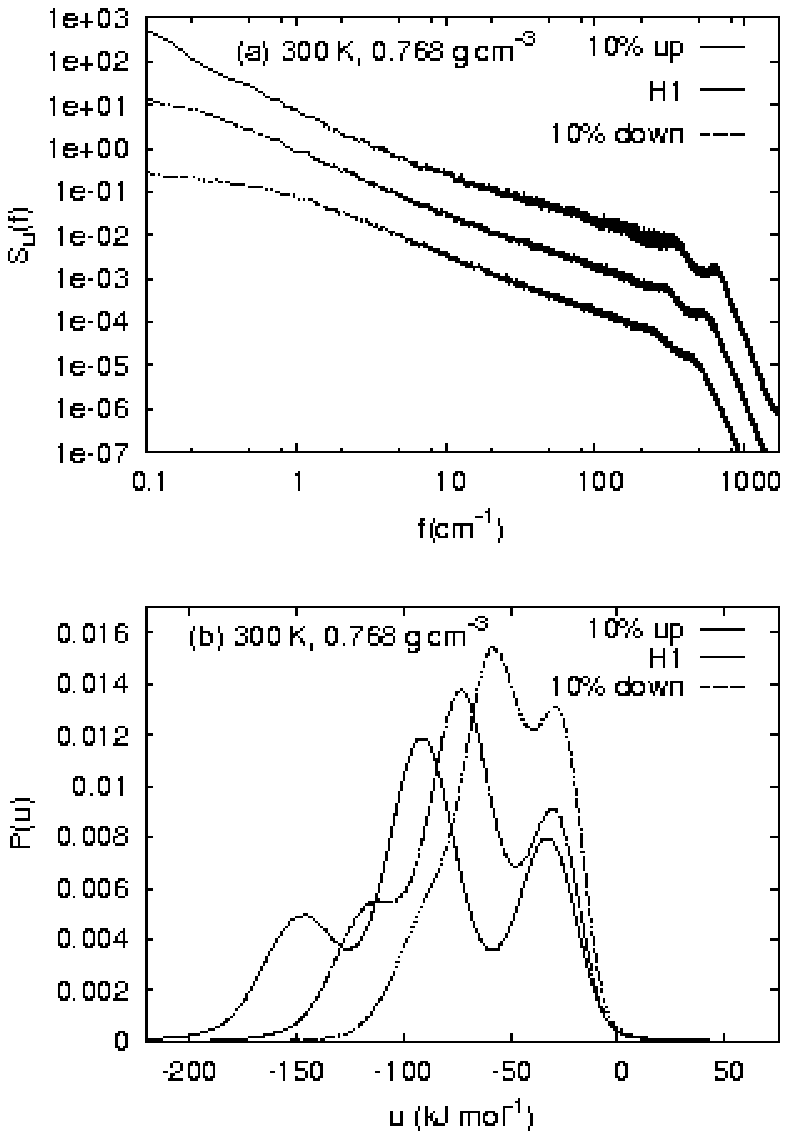}
\end{figure}

\newpage
\begin{figure}
\caption{\quad}
\includegraphics[trim=1.5truein 0 0 2truein]{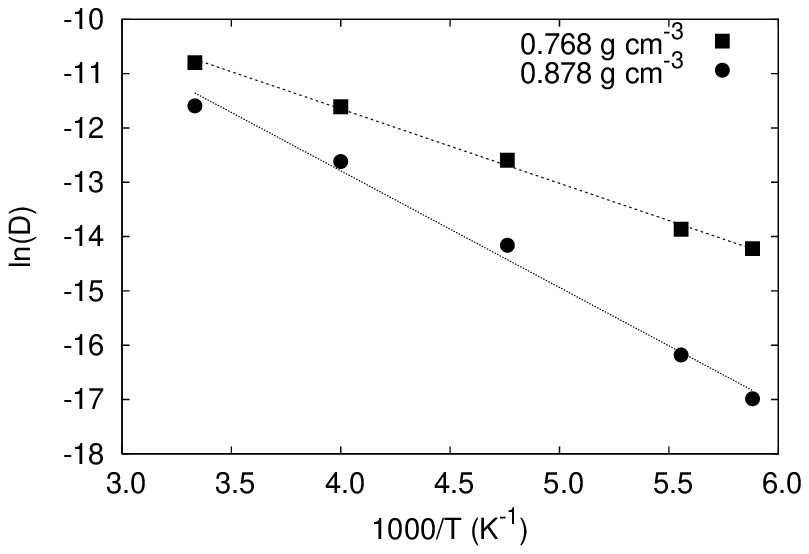}
\end{figure}

\newpage
\begin{figure}
\caption{\quad}
\includegraphics[trim=1.5truein 0 0 2truein]{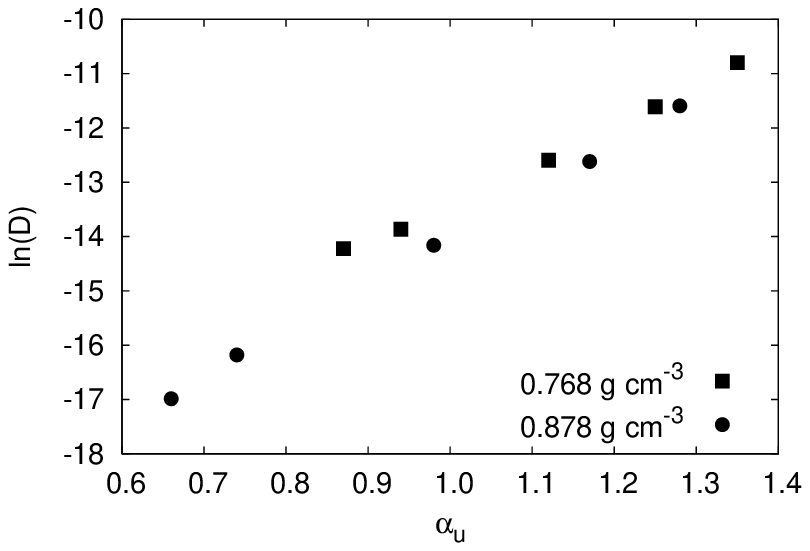}
\end{figure}

\newpage
\begin{figure}
\caption{\quad}
\includegraphics[trim=1.5truein 0 0 2truein]{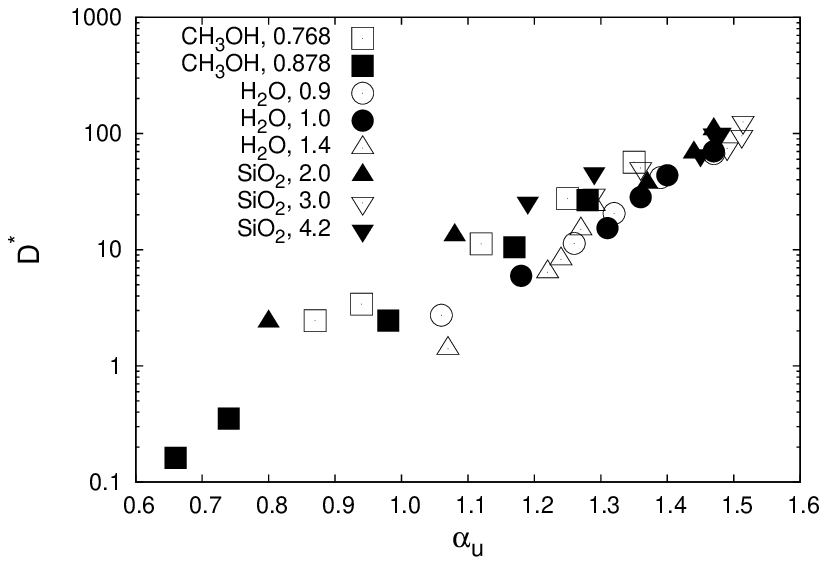}
\end{figure}


\begin{thebibliography}{99}

\bibitem{ls93} Ladanyi, B. M.; Skaf M.S.
\arpc{1993}{44}{335}

\bibitem{bb03} Bagchi, B.
\arpc{2003}{99}{127}

\bibitem{bb05} Bagchi, B.
\chemrev{2005}{105}{3197}

\bibitem{mcbf99} {\it Hydration Processes in Biology} edited by Bellissent-Funel, M.-C. (IOS Press, Amsterdam, 1999)

\bibitem{lrt061} Loparo, J. L.; Roberts, S. T.; Tokmakoff, A.
\jcp{2006}{125}{194521}

\bibitem{lrt062} Loparo, J. L.; Roberts, S. T.; Tokmakoff, A.
\jcp{2006}{125}{194522}

\bibitem{rb05} Rezus, Y. L. A.; Bakker, H. J.
\jcp{2005}{123}{114502}

\bibitem{rlp04} Raiteri, P.; Laio, A.; Parrinello, M.
\prl{2004}{93}{087801}

\bibitem{lc96} Luzar, A.; Chandler, D.
\prl{1996}{76}{928}

\bibitem{at07} Tokmakoff, A.
\science{2007}{317}{54}

\bibitem{jt99} Texeira, J. in {\it Hydration Processes in Biology} edited by Bellissent-Funel, M.-C. (IOS Press, Amsterdam, 1999)

\bibitem{wlj86} Jorgensen, W. L.
\jpc{1986}{90}{1276}
 
\bibitem{yhs99} Yamaguchi, T.; Hidaka, K.; Soper, A. K.
\mp{1999}{96}{1199}; erratum \mp {1999}{97}{603}

\bibitem{ybs00} Yamaguchi, T.; Benmore, C. J.; Soper, A.K.
\jcp{2000}{112}{8976}

\bibitem{wnfz00} Weitkamp, T.; Neuefeind, J.; Fischer, H. E.; Zeidler, M. D.
\mp{2000}{98}{125}

\bibitem{mw00}Madden, P. A. and Wilson, M.\jpcm{12}{A95}{2000} 

\bibitem{wu45} Wang, M. C.; Uhlenbeck, G. E. {\it Rev. Mod. Phys.} {\bf 1945} {\it 17}, 323.

\bibitem{mrc03} Mudi, A.; Ramaswamy, R.; Chakravarty, C.
\cpl{2003}{376}{683}

\bibitem{mc04} Mudi, A.; Chakravarty, C.
\jpcb{2004}{108}{19607}

\bibitem{mc05} Mudi, A.; Chakravarty, C.
\jcp{2005}{122}{104507}

\bibitem{mc06} Mudi, A.; Chakravarty, C.
\jpcb{2006}{110}{8422}

\bibitem{mcm06} Mudi, A.; Chakravarty, C.; Milotti, E.
\jcp{2006}{125}{074508}

\bibitem{smc06} Sharma, R.; Mudi, A.; Chakravarty, C.
\jcp{2006}{125}{044705}


\bibitem{sor92} Sasai, M.; Ohmine, I.; Ramaswamy, R.
\jcp{1992}{96}{3045}

\bibitem{hfm87} Haughney, M.; Ferrario M.; McDonald, I. R.
\jpc{1987}{91}{4934}

\bibitem{wpgy96} Wallen, S.L.; Palmer, B.J.; Garrett, B. C.; Yonker, C. R.
\jpc{1996}{100}{3959}

\bibitem{an98} Asahi, N.; Nakamura, Y.
\jcp{1998}{109}{9879}

\bibitem{vcs00} Venables, D. S.; Chiu, A.; Schmuttenmaer C. A.
\jcp{2000}{113}{3243}

\bibitem{tkt99} Tsuchida, E.; Kanada, Y.; Tsukada, M.
\cpl{1999}{311}{236}

\bibitem{hmg04} Handgraaf, J-W.; Meijer, E. J.; Gaigeot, M-P.
\jcp{2004}{121}{10111}

\bibitem{sf96} Smith, W.; Forester, T. R.
{\it J. Mol. Graphics} {\bf 1996}, {\it 14}, 136.
                                                                                
\bibitem{syr01} Smith, W.; Yong, C. W.; and Rodger, P. M.
\molsim{2002}{28}{385};
 The DL\_POLY website is {\it http://www.cse.clrc.ac.uk/msi/software/DL\_POLY/}.

\bibitem{sk92} Sindzingre, P.; Klein, M. L.
\jcp{1992}{96}{4681}

\bibitem{mm03} Martinez, J. M.; Martinez, L.
\jcompphys{2003}{24}{819}

\bibitem{hm86} Hansen, J.-P.; McDonald, I. R.; {\it Theory of
Simple Liquids} (Academic Press, 1986).

\bibitem{numrec} Press, W. H.; Flannery, B. P.; Teukolsky, S. A.; Vetterling, W. T. : {\it Numerical Recipes in Fortran} (Cambridge University Press, Cambridge, 1990).

\bibitem{mi86} Madden, P. A.; Impey, R. W. \cpl{123}{502}{1986}

\bibitem{bg91} Guillot, B. \jcp{95}{1543}{1991}

\bibitem{ph98} Parker, M. E.; Heyes, D. M. \jcp{108}{9039}{1998}

\bibitem{pkm70} Passchier, W. F.; Klompmaker, E. R.; Mandel, M.
\cpl{1970}{4}{485}

\bibitem{gmo90} Guillot, B.; Marteau, P.; Obriot, J.
\jcp{1990}{93}{6148}

\bibitem{mg90} Matsumoto, M; Gubbins, K. E. \jcp{93}{1981}{1990}

\bibitem{cccrp99} Chelli, R.; Ciabatti, S.; Cardini, G., Righini, R.; Procacci,
P.
\jcp{1999}{111}{4218}

\bibitem{pcrs03} Pagliai, M.; Cardini, G., Righini, R.; Schettino V.
\jcp{2003}{119}{6655}





















\end{thebibliography}
\end{document}